\documentclass[aps,prx,amsmath,amssymb,amsfonts,10pt,english,twocolumn,superscriptaddress,nofootinbib,floatfix,showkeys]{revtex4-2}
\usepackage{graphicx}
\usepackage{subfigure}
\usepackage{amsmath, amsthm, amssymb, amsfonts}
\usepackage{verbatim}
\usepackage{dcolumn}
\usepackage{bm}
\usepackage{epsf}
\usepackage{color}
\usepackage[colorlinks=true,citecolor=blue,linkcolor=blue,urlcolor=blue]{hyperref}%
\usepackage{xcolor}
\usepackage{physics}
\usepackage{dsfont}
\usepackage{tikz}
\usepackage{multirow}
\usepackage{xfrac}
\usepackage{makecell}
\usepackage[normalem]{ulem}
\usepackage{siunitx}
\usepackage{rotating}
\usepackage{bbm}

    \DeclareMathOperator*{\argmin}{arg\,min}

    \newcommand{\sigx}{\sigma^x}
    
    \newcommand{\sigz}{\sigma^z}

\newcommand{\etal}{\textit{et al. }}

\newcommand{\bla}{bla\\bla\\bla\\bla\\bla}

\DeclareMathAlphabet\mathbfcal{OMS}{cmsy}{b}{n}

\makeatletter
\newcommand{\currentfontsize}{The current font size is: \f@size pt}
\newcommand{\currenttextwidth}{The current text width is: \the\textwidth}
\newcommand{\currentcolumnwidth}{The current column width is: \the\columnwidth}
\makeatother

\begin{document}

\title{Variational Gibbs State Preparation on Trapped-Ion Devices}

\author{Reece Robertson}
\email{reecerobertson@umbc.edu}
\affiliation{Department of Computer Science and Electrical Engineering, University of Maryland, Baltimore County, Baltimore, MD 21250, USA}
\affiliation{Department of Physics, University of Maryland, Baltimore County, Baltimore, MD 21250, USA}
\affiliation{Quantum Science Institute, University of Maryland, Baltimore County, Baltimore, MD 21250, USA}

\author{Mirko Consiglio}
\affiliation{Department of Physics, University of Malta, Msida MSD 2080, Malta}

\author{Josey Stevens}
\affiliation{Department of Physics, University of Maryland, Baltimore County, Baltimore, MD 21250, USA}
\affiliation{Quantum Science Institute, University of Maryland, Baltimore County, Baltimore, MD 21250, USA}

\author{Emery Doucet}
\affiliation{Department of Physics, University of Massachusetts, Boston, Boston, MA 02125, USA}

\author{Tony J. G. Apollaro}
\affiliation{Department of Physics, University of Malta, Msida MSD 2080, Malta}

\author{Sebastian Deffner}
\affiliation{Department of Physics, University of Maryland, Baltimore County, Baltimore, MD 21250, USA}
\affiliation{Quantum Science Institute, University of Maryland, Baltimore County, Baltimore, MD 21250, USA}
\affiliation{National Quantum Laboratory, College Park, MD 20740, USA}

\begin{abstract}
    We implement a variational quantum algorithm for Gibbs state preparation of a transverse-field Ising model on IonQ's quantum computers.
    To this end, we train the variational parameters via classical simulation and perform state tomography on the quantum devices to evaluate the fidelity of the prepared Gibbs state.
    As a main result, we find that fidelity decreases (non-monotonically) as a function of the inverse temperature $\beta$ of the system.
    Fidelity also decreases as a function of the size of the system. Interestingly, we find that a Gibbs state prepared for a specified $\beta$ is a better representative of a Gibbs state prepared for a \textit{lower} $\beta$; or in other words, thermal fluctuations in the quantum hardware lead to digital heating, that is, an increase in the temperature of the prepared Gibbs state above what was intended.
\end{abstract}

\maketitle

\section{Introduction}\label{sec:introduction}

From its very inception, quantum computing has been envisioned as a tool to numerically model complex quantum systems \cite{feynman_simulating_1982}.
Complex systems, however, are usually best described by notions of thermodynamics and statistical physics, and hence preparing Gibbs states on quantum computers is an important problem.
See, for instance, recent reviews of hybrid algorithms \cite{endo_hybrid_2021}, quantum algorithms with super-quadratic speedup \cite{babbush_grand_2025, dalzell_quantum_2023}, and scalable noisy intermediate-scale quantum (NISQ) algorithms \cite{bharti_noisy_2022, gacon_scalable_2024}.

From a physics perspective, Gibbs states are interesting both in their own right and in what they reveal about the Hamiltonians from which they arise.
However, from the point of view of algorithm development, it is interesting to prepare Gibbs states on quantum computers.
For instance, many quantum machine learning algorithms use Gibbs states as an initial state for a Metropolis-Hastings sampling routine.
This includes the quantum Boltzmann machine framework, which requires a Gibbs initial state \cite{shingu_boltzmann_2021, zoufal_variational_2021}; with applications in machine learning tasks such as quantum reinforcement learning \cite{jerbi_quantum_2021, gerlach_quantum_2025}, supervised learning \cite{schuld_learning_2018}, structured prediction \cite{sepehry_smooth_2019}, Markov chain Monte Carlo \cite{wittek_quantum_2017}, and neural networks \cite{hagai_artificial_2023}.
More broadly, Gibbs state preparation (GSP) falls within the general class of hybrid quantum--classical algorithms for quantum state preparation \cite{avramouli_quantum_2022, anshu_survey_2024}.

While quantum machine learning receives much of the attention regarding GSP on quantum devices, there are several other significant applications.
In quantum thermodynamics, Gibbs states are used to model open system dynamics \cite{haug_generalized_2022}, they enable efficient routines for quantum state tomography \cite{gupta_variational_2022, lifshitz_practical_2023}, and ground state preparation \cite{wang_ground_2023}.
In addition, Gibbs states are employed to compute the trace distance (fidelity) \cite{chen_variational_2021}, maximum likelihood \cite{piatkowski_quantum_2024}, and matrix exponentials \cite{subramanian_quantum_2021}.
Finally, there are also applications in finite temperature quantum chemistry \cite{bidart_quantum_2025, bauer_efficient_2025}, and computer science \cite{dalzell_quantum_2023}.

Many GSP algorithms operate as variational quantum algorithms \cite{chowdhury_variational_2020, jerbi_quantum_2021, sewell_thermal_2022, warren_adaptive_2022, gao_adaptive_2023, huijgen_training_2024, gu_practical_2024, rouze_optimal_2024, smid_polynomial-time_2025}, with, e.g., imaginary time evolution \cite{zoufal_variational_2021, turro_quantum_2023}, cluster expansions \cite{eassa_gibbs_2024}, and parallelizable local measurements \cite{chen_learning_2025}, leveraging hardware noise \cite{foldager_noise-assisted_2022, dambal_noise-induced_2025}, 
or simultaneous optimization of multiple targets \cite{tuan_hai_multi-target_2024}.
In addition,  \citet{magann_randomized_2023} developed a GSP algorithm which is not variational in the traditional sense (it does not employ a classical optimization routine), but rather operates using randomized circuits with measurements taken in random directions in the tangent space.
\citet{economou_role_2023} studied the role of entanglement in these algorithms, and \citet{benedetti_variational_2021} studied quantum variational inference methods without specifically applying their techniques to Gibbs sampling.

In our work, we use the VQA developed by Consiglio \etal \cite{consiglio_variational_2024, consiglio_thermal_2024, consiglio_variational_2025}, which was successfully implemented on superconducting hardware \cite{consiglio_variational_2024, edo_study_2025, morstedt_rapid_2025}.
Interestingly, developments for trapped-ion hardware are much more nascent; a theoretical proposal was made in 2021 \cite{reiter_engineering_2021}, and an implementation on Quantinuum hardware was demonstrated in 2025 \cite{granet_adiabatic_2025}.
Yet, due to their full connectivity, an implementation in trapped ions is desirable since the GSP algorithm does not map natively onto the heavy-hex topology used by IBM devices; \texttt{SWAP} operations (which are particularly noisy \cite{robertson_simons_2025}) must be inserted into the circuit during the compilation step.
Running the algorithm on fully-connected trapped-ion machines obviates all difficulties caused by topology and allows for a more compact representation of the algorithm.\footnote{In that connection, the GSP algorithm can also be mapped directly onto a sufficiently large square grid topology (like those used by the Rigetti superconducting devices) without the addition of \texttt{SWAP} operations.}
In the present work, we report the preparation of Gibbs states for a transverse-field Ising model (TFIM) at various sizes on IonQ's trapped-ion quantum computers. While generally successful, we find that hardware noise decreases fidelity as the temperature decreases.

In this paper, we describe our experimental results from preparing and validating Gibbs states on IonQ trapped-ion hardware.
We describe our algorithm and approach in the next section, and report on the results of our experiments in Section \ref{sec:results}.
Section \ref{sec:conclusion} summarizes our findings, and additional details on the hardware and algorithm are presented in Appendices \ref{sec:hardware} \& \ref{sec:example}, respectively.

\section{Methods}\label{sec:methods}

We begin by outlining the details of our method.
We use a hybrid quantum--classical algorithm, for which we discuss both the quantum and classical aspects.

As already mentioned, we employ the procedure introduced in \citet{consiglio_variational_2024}.
We consider a Hamiltonian, $\mathcal{H}$, describing $n$ qubits with eigenvalues $E_i$ and eigenvectors $|E_i\rangle$.
Then the corresponding Gibbs state reads,
\begin{equation}
    \rho(\beta,\mathcal{H}) = \frac{e^{-\beta\mathcal{H}}}{\mathcal{Z}(\beta,\mathcal{H})},
    \label{eq:gibbs}
\end{equation}
where $\beta = 1/k_BT$ is the inverse temperature and, $\mathcal{Z}(\beta,\mathcal{H}) = \text{tr}\left\{e^{-\beta\mathcal{H}}\right\}$ is the partition function. As usual \cite{Callen1985} the Helmholtz free energy then is
\begin{equation}
    \mathcal{F}(\rho) = \text{tr}\left\{\mathcal{H}\rho\right\} - \beta^{-1} \mathcal{S}(\rho),
    \label{eq:free-energy}
\end{equation}
where $\mathcal{S}(\rho)$ is the von Neumann entropy,
\begin{equation}
    \mathcal{S}(\rho) = -k_B \sum_{i=0}^{2^n-1} p_i\ln{p_i},
    \label{eq:von-Neumann}
\end{equation}
and $p_i$ are the eigenvalues of $\rho(\beta,\mathcal{H})$.
Now recalling that the Gibbs state minimizes the free energy of $\mathcal{H}$ \cite{Callen1985}, we compute Eq.~\eqref{eq:gibbs} through the optimization function
\begin{equation}
    \rho(\beta,\mathcal{H}) = \argmin_\rho \mathcal{F}(\rho).
    \label{eq:optimization}
\end{equation}
For the present purposes, we implement $\mathcal{F}(\rho)$ through a variational GSP algorithm. 

\begin{figure}[t]
    \centering
    \includegraphics[width=0.6\linewidth]{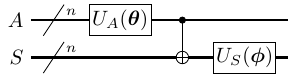}
    \caption{
        An overview of $U_G(\boldsymbol\theta,\boldsymbol\phi)$, the parametrized quantum algorithm for Gibbs state preparation.
        This algorithm acts on two $n$-qubit quantum registers, an ancilla register $A$ and a system register $S$.
        The algorithm consists of two unitaries: $U_A(\boldsymbol\theta)$, which acts on the ancilla register, and $U_S(\boldsymbol\phi)$, which acts on the system register.
        Between these two unitaries, a layer of \texttt{CNOT} gates is applied transversely between the registers.
    }
    \label{fig:alg-overview}
\end{figure}

In the following, we refer to the parameterized unitary that implements the GSP algorithm on a quantum device as $U_G(\boldsymbol\theta,\boldsymbol\phi)$.
An overview of $U_G(\boldsymbol\theta,\boldsymbol\phi)$ is depicted in Fig.~\ref{fig:alg-overview}.
This algorithm operates on two quantum registers: an $n$-qubit ancilla register $A$ and an $n$-qubit system register $S$.
A parameterized unitary $U_A(\boldsymbol\theta)$ is applied to the ancilla register to enact a probability distribution preparation protocol.
The result is transmitted to the system register via a layer of transversal \texttt{CNOT} gates.
That is, a \texttt{CNOT} gate conditioned on qubit $a_i$ with target qubit $s_i$, where $\{a_i:i\in[1,n]\}$ are the qubits of $A$ and $\{s_i:i\in[1,n]\}$ are the qubits of $S$.
In combination, these unitaries initialize the system register in a non-trivial probability distribution over the computational basis states.
At this point, $U_S(\boldsymbol\phi)$ acts on the system register.
This unitary is parametrized such that it transforms the basis states of the prepared probability distribution from the computational basis into the eigenbasis of the Hamiltonian of interest.

\begin{figure}[t]
    \centering
    \includegraphics[width=0.9\linewidth]{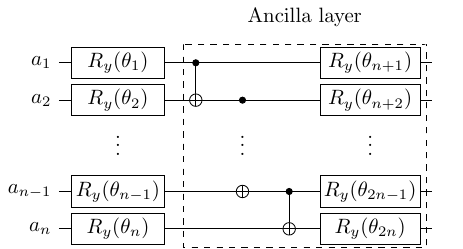}
    \caption{
        The details of $U_A(\boldsymbol\theta)$ (see Fig.~\ref{fig:alg-overview}), the state-preparation unitary for the ancilla register.
        This unitary consists of two columns of $R_y(\theta_i)$ gates for $i\in[1,n]$, separated by a layer of \texttt{CNOT} gates acting between adjacent qubits $a_i$ and $a_{i+1}$ for $i\in[1,n-1]$.
        The gates within the dashed box constitute one \textit{ancilla layer}.
        In general, one can perform an arbitrary number of ancilla layers.
        In this paper, we consistently use one ancilla layer in all experiments.
    }
    \label{fig:alg-ancilla}
\end{figure}

The structure of the unitary $U_A(\boldsymbol\theta)$ is depicted in Fig.~\ref{fig:alg-ancilla}.
Observe that it consists of a transversal layer of $R_y(\theta_i)$ gates for $i\in[1,n]$ (where $\theta_i\in\boldsymbol\theta$ for all $i\in[1,2n]$).
Following this, there is a layer of \texttt{CNOT} gates acting on qubits $a_i$ and $a_{i+1}$ for $i\in[1,n-1]$,
and then another layer of $R_y(\theta_i)$ gates for $i\in[n+1,2n]$.
Note that the last two layers of operations constitute one \textit{ancilla layer}; this ancilla layer can be repeated arbitrarily many times before the transversal \texttt{CNOT} mapping between the ancilla and system registers.
In all of our experiments, we use exactly one ancilla layer.

\begin{figure*}[t]
    \centering
    \includegraphics[width=.9\textwidth]{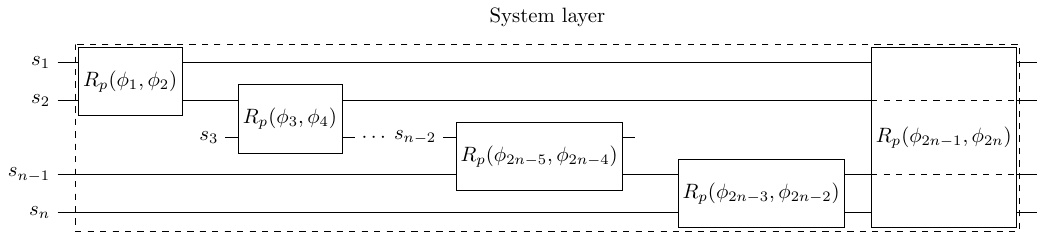}
    \caption{
        The implementation of the $U_S(\boldsymbol\phi)$ of Fig.~\ref{fig:alg-overview}.
        This operation consists of the application of an $R_p(\theta_{2i-1},\theta_{2i})$ gate to qubits $s_i$, $s_{i+1}$ (where $s_{n+1}=s_1$) for $i\in[1,n]$.
        For the implementation of the $R_p(\theta_{2i-1},\theta_{2i})$ gate, see Fig.~\ref{fig:alg-rp}.
        The collection of $n$ $R_p(\theta_{2i-1},\theta_{2i})$ gates constitutes a \textit{system layer}.
        In general, multiple system layers can be performed within $U_G(\boldsymbol\theta,\boldsymbol\phi)$; in this paper, we consistently use one system layer.
    }
    \label{fig:alg-system}
\end{figure*}

\begin{figure*}[t]
    \centering
    \includegraphics[width=.9\textwidth]{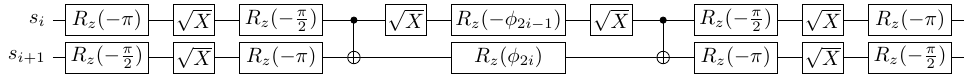}
    \caption{
        The definition of the $R_p(\phi_{2i-1},\phi_{2i})$ gate.
        Note that this gate operates on qubits $s_i$ and $s_{i+1}$.
        Moreover, this is an abstract representation of the operation; for execution on hardware, this must be transpiled into hardware native gates.
        For more on hardware native gates, see Appendix \ref{sec:hardware}.
    }
    \label{fig:alg-rp}
\end{figure*}

Similarly, the structure of $U_S(\boldsymbol\phi)$ is given in Fig.~\ref{fig:alg-system}.
Here, one \textit{system layer} constitutes a cyclic application of $R_p(\phi_{2i-1},\phi_{2i})$ gates to neighboring qubits, where qubit $s_1$ and qubit $s_n$ are considered neighbors (hence the cyclic structure of this operation).
Explicitly, the $i^\text{th}$ $R_p(\theta_{2i-1},\theta_{2i})$ gate acts on qubits $s_i$ and $s_{i+1}$, where $s_{n+1}=s_1$.
As before, each $\phi_{2i-1}, \phi_{2i} \in \boldsymbol\phi$ for all $i\in[1,n]$.
Similar to the ancilla layer, the system layer can be repeated an arbitrary number of times before the final measurement, and again, we consistently apply one system layer in our experiments.
The structure of the $R_p(\phi_{2i-1},\phi_{2i})$ is shown in Fig.~\ref{fig:alg-rp}.

A complete example for $n=4$ using one ancilla layer and one system layer is shown in Fig.~\ref{fig:alg-example} of Appendix \ref{sec:example}.
This circuit represents the most complex GSP algorithm executed on quantum hardware during the course of this study.

In terms of $U_G(\boldsymbol\theta, \boldsymbol\phi)$,  Eq.~\eqref{eq:optimization} then reads,
\begin{equation}
    \begin{aligned}
        \rho(\beta,\mathcal{H}) &= \argmin_{\boldsymbol\theta,\boldsymbol\phi} \mathcal{F}(\rho(\boldsymbol\theta,\boldsymbol\phi))\\
        &= \argmin_{\boldsymbol\theta,\boldsymbol\phi} \left[\text{tr}\left\{\mathcal{H}\rho_S(\boldsymbol\theta,\boldsymbol\phi)\right\}-\beta^{-1}\mathcal{S}(\rho_A(\boldsymbol\theta))\right],
    \end{aligned}
    \label{eq:cost-function}
\end{equation}
where $\rho_S(\boldsymbol\theta, \boldsymbol\phi)$ is represented by the action of $U_G(\boldsymbol\theta, \boldsymbol\phi)$ on the initial state with the ancilla register traced out, while $\rho_A(\boldsymbol\theta)$ represents the state of the ancilla register just after applying $U_A(\boldsymbol\theta)$ and the layer of \texttt{CNOT} gates.

Now, we turn our attention to the classical preprocessing, optimization loop, and postprocessing necessary to execute and validate the quantum GSP algorithm.
For clarity, we use the term \textit{GSP algorithm} to specifically refer to the quantum circuit of Fig.~\ref{fig:alg-overview}, and we use the term \textit{GSP routine} to refer to the entire end-to-end protocol.
This includes all of the classical computation as well as the quantum GSP algorithm.

To begin the GSP routine, we must select a Hamiltonian $\mathcal{H}$.
In this work, we chose to work with the transverse-field Ising model (TFIM),
\begin{equation}
    \mathcal{H} = -\frac{1}{2} \sum_{i=1}^{n}\sigx_i\sigx_{i+1} - h\sum_{i=1}^n \sigz_i,
    \label{eq:ising-hamiltonian}
\end{equation}
with periodic boundary conditions, i.e., $n + 1 = 1$.

For each set of values of $n$, $h$, and $\beta$, the vectors $\boldsymbol\theta,\boldsymbol\phi\in\mathbb{R}^{2n}$ were initialized with randomly selected elements in the interval $[-\pi,\pi]$.
Throughout the training protocol, the elements of these vectors were permitted to take on any real value.

The initialization of $\boldsymbol\theta$ and $\boldsymbol\phi$ completes the preprocessing step of the GSP routine.
The next step is to optimize $\boldsymbol\theta$ and $\boldsymbol\phi$ by evaluating Eq.~\eqref{eq:cost-function}.\footnote{To keep costs reasonable in the era of NISQ computing, we performed this step via classical simulation.}
We elected to use the Simultaneous Perturbation Stochastic Approximation (SPSA) optimizer implemented by \texttt{Qiskit} for the minimization problem \cite{qiskit_docs}.
Moreover, we fixed the maximum number of optimization iterations at 100.

The $i^\text{th}$ iteration of the optimization loop proceeded as follows:
\begin{enumerate}
    \item The $i^\text{th}$ instance of the quantum GSP algorithm, $U_G\left(\boldsymbol\theta^{(i)},\boldsymbol\phi^{(i)}\right)$, is generated.
    \item 16,384 shots of $U_G\left(\boldsymbol\theta^{(i)},\boldsymbol\phi^{(i)}\right)$ are evaluated on a classical noisy quantum simulator (on both registers).
    \begin{enumerate}
        \item 8,192 shots measured the system register in the $\sigz$ basis.
        \item 8,192 shots measured the system register in the $\sigx$ basis.
    \end{enumerate}
    \item Each of the $\expval{\sigx_i\sigx_{i+1}}_{\rho_S}$ expectation values for $i \in [1,n]$ are evaluated from the $\sigx$ basis shots since they are commuting terms.
    \item Likewise, each of the $\expval{\sigz_i}_{\rho_S}$ expectation values for $i \in [1,n]$ are evaluated from the $\sigz$ basis shots since they are commuting terms.
    \item The von Neumann entropy $\mathcal{S}\left(\rho_A\left(\boldsymbol\theta^{(i)}\right)\right)$
     is computed from the measurement results of all 16,384 shots in the $\sigz$ basis, since the diagonal terms of $\rho_A$ are the eigenvalues $p_i$ of $\rho_S$ \cite{consiglio_variational_2024}.
    \item The cost function, Eq. \eqref{eq:cost-function}, is computed and used by the SPSA optimizer to generate $\boldsymbol\theta^{(i+1)}$ and $\boldsymbol\phi^{(i+1)}$.
\end{enumerate}

At the conclusion of the optimization protocol, the optimal parameters $\boldsymbol\theta$ and $\boldsymbol\phi$ are used to generate the final circuit $U_G(\boldsymbol\theta,\boldsymbol\phi)$.
This circuit is then evaluated on \textit{quantum} hardware to prepare a Gibbs state.
The quality of the GSP algorithm is verified by performing state tomography on the quantum device to reconstruct the final system density matrix.
This density matrix is compared to the analytical Gibbs state of $\mathcal{H}$, computed via exact diagonalization.

\subparagraph{Employed hardware}

Before moving on to the results, we conclude this section by briefly discussing the quantum hardware used in this study.
All experiments were executed on IonQ trapped-ion devices.
Three devices were used: the 25-qubit IonQ Aria 1, the 36-qubit IonQ Forte, and the 36-qubit IonQ Forte Enterprise.
All devices are fully-connected, and the important hardware parameters are reported in Appendix \ref{sec:hardware}.

\begin{figure}[t]
    \centering
    \includegraphics[width=\linewidth]{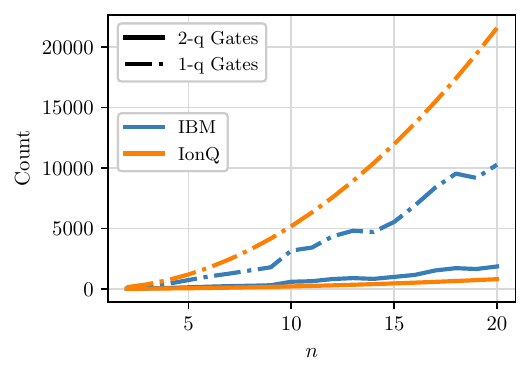}
    \caption{
        The number of one- and two-qubit native gates required to implement the GSP algorithm on both IBM hardware and IonQ hardware.
        All native-gate circuits were generated using the Qiskit transpiler, targeting the Brisbane and Aria backends from IBM and IonQ, respectively.
        Note that while the number of one-qubit gates scales more rapidly on IonQ hardware, the opposite is true for two-qubit gates.
        Since the dominant source of gate error arises from two-qubit gates, it follows that the IonQ hardware potentially outperforms the IBM hardware in terms of algorithm fidelity.
    }
    \label{fig:gate-counts}
\end{figure}

As mentioned above, an important advantage of trapped-ion hardware in executing the GSP algorithm is all-to-all qubit connectivity.
This allows for an implementation of the algorithm without the use of \texttt{SWAP} operations, which leads to lower two-qubit gate counts as compared to superconducting hardware (see Fig.~\ref{fig:gate-counts}).
The dominant source of gate error on NISQ hardware stems from two-qubit gates \cite{ionq_noise}, and \texttt{SWAP} gates are particularly noisy \cite{robertson_simons_2025}.
Note that expressing the GSP routine in IonQ hardware native gates does yield a larger total gate count than the same circuits expressed in IBM native gates.
However, since this increase in total gate count is due to high-fidelity one-qubit gates, this impact is outweighed by the impact of the two-qubit gates within the circuit.

\section{Results}\label{sec:results}

For our hardware runs, we chose the following values of the parameters: $n\in\{2,3,4\}$, $h\in\{0.5,1.0,1.5\}$, and $\beta\in\{10^{-8},1,5\}$.
Note that we use $10^{-8}$ as a close proxy for the $\beta=0$ infinite temperature case.
In total, we evaluated 1,701 circuits on hardware for a cost of \$166,917.86 USD in IonQ compute credits.

\begin{figure}[tbp]
    \centering
    \includegraphics[width=\linewidth]{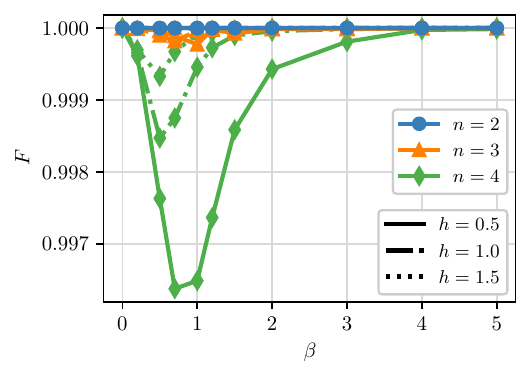}
    \caption{
        Results from simulating the entire GSP routine on a noiseless simulator.
        Observe that fidelity is high for the extreme values of $\beta$, and that it reaches its minimum near $\beta \sim 1$.
        Moreover, we see that fidelity decreases as $n$ increases, and that fidelity increases as $h$ increases.
    }
    \label{fig:simulated-fidelities}
\end{figure}

To begin, we classically simulated our entire GSP routine on a noiseless statevector simulator.
We repeated this simulation 100 times and recorded the highest fidelity achieved for each parameter configuration.
The results are reported in Fig.~\ref{fig:simulated-fidelities}.
From this figure, we see that fidelity is near unity for $\beta=0$ and $\beta=5$, and that fidelity decreases for values of $\beta$ between these regions, with the worst observed performance occurring in the regime $\beta \sim 1$.
One subtlety to note is that the observed minimum in the figure likely does not coincide with the minimum of the function; rather, we are sampling points from this fidelity function and interpolating between them.
That said, the interpolation suggests the region in which the function will attain its minimum.

Moreover, from Fig.~\ref{fig:simulated-fidelities}, we see that increasing $n$ decreases algorithm fidelity, while increasing $h$ increases fidelity.
These results are intuitive; adding more qubits increases noise within the system, which decreases fidelity.
Increasing $h$, on the other hand, increases the strength of the ferromagnetic field, which suppresses local system fluctuations.
In algorithmic terms, increasing $h$ increases the weight of the higher-fidelity single-qubit rotations in Eq.~\eqref{eq:ising-hamiltonian} over the lower-fidelity two-qubit rotations, leading to an improved overall fidelity.

\begin{figure*}[p]
    \centering
    \includegraphics[width=\linewidth]{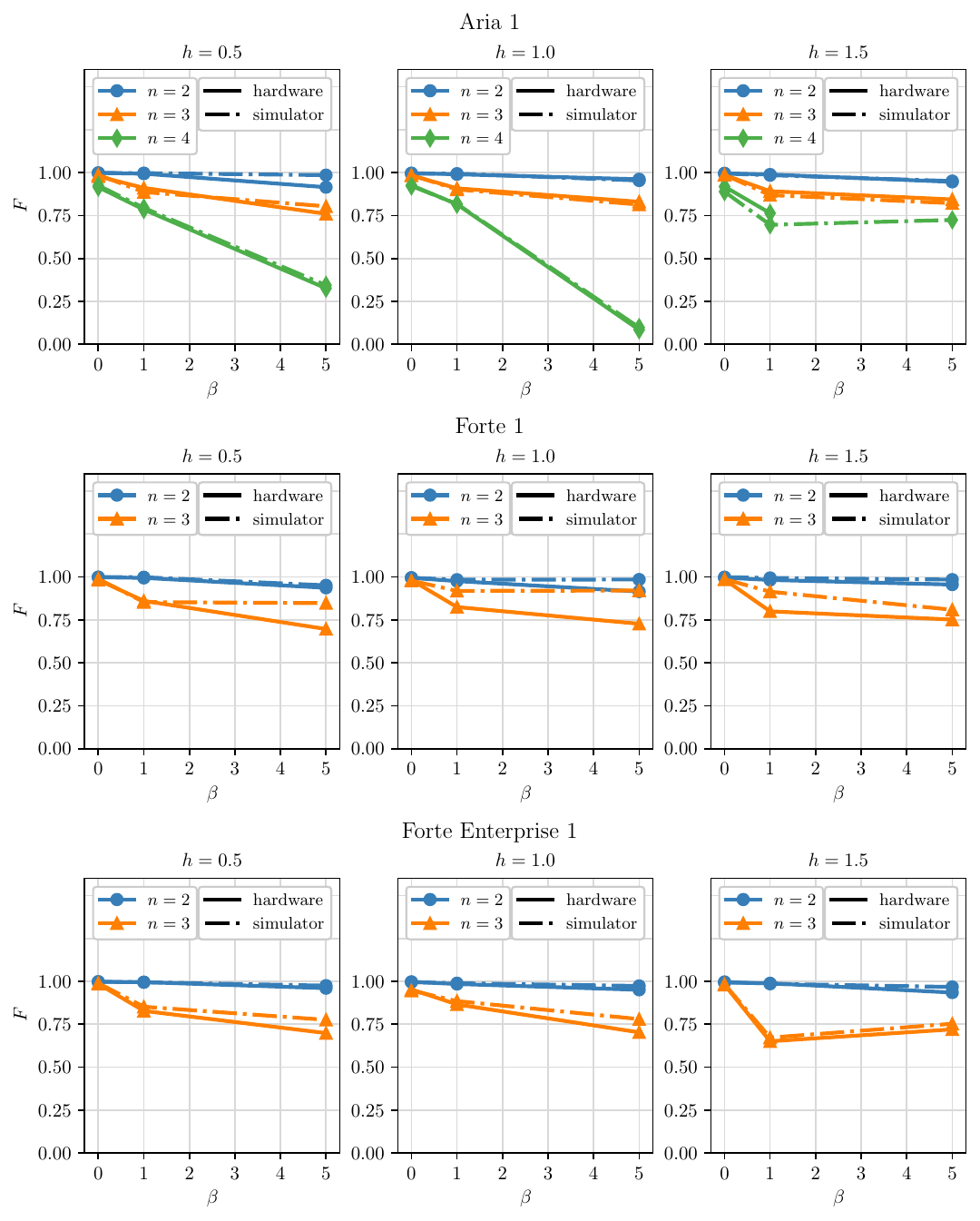}
    \caption{
        The results of our GSP routine as executed on IonQ hardware and noisy simulators.
        Increasing $n$ decreases the algorithm fidelity, which is consistent with the noiseless simulator results.
        However, due to hardware noise, increasing $h$ does not consistently improve fidelity, and fidelity is not near unity for $\beta=5$.
        The top right plot is missing a $\beta=5$ point as Aria 1 went offline while the job was queued.
        The $n=4$ data are missing for Forte 1 and Forte Enterprise 1 for the same reason.
    }
    \label{fig:fidelities}
\end{figure*}

Armed with intuition garnered from our noiseless simulations, we proceed with the results from the IonQ hardware.
To obtain these results, we classically optimized the VQA for each hardware device and value of $n$, $h$, and $\beta$.
We then ran the circuits on the real hardware device and performed tomography on the system register in order to reconstruct the density matrix of the system.
We performed 1,024 shots for each observable measured during the state tomography routine.
We finally computed the exact Gibbs state for each combination of variables classically and compared the fidelity \cite{uhlmann_transition_2011}, $F(\rho,\sigma)=\left(\text{tr}\left\{\sqrt{\sqrt{\rho}\sigma\sqrt{\rho}}\right\}\right)^2$, between the experimental and exact results.
The results are collected in Fig.~\ref{fig:fidelities}.

\begin{figure*}[p]
    \centering
    \includegraphics[width=\linewidth]{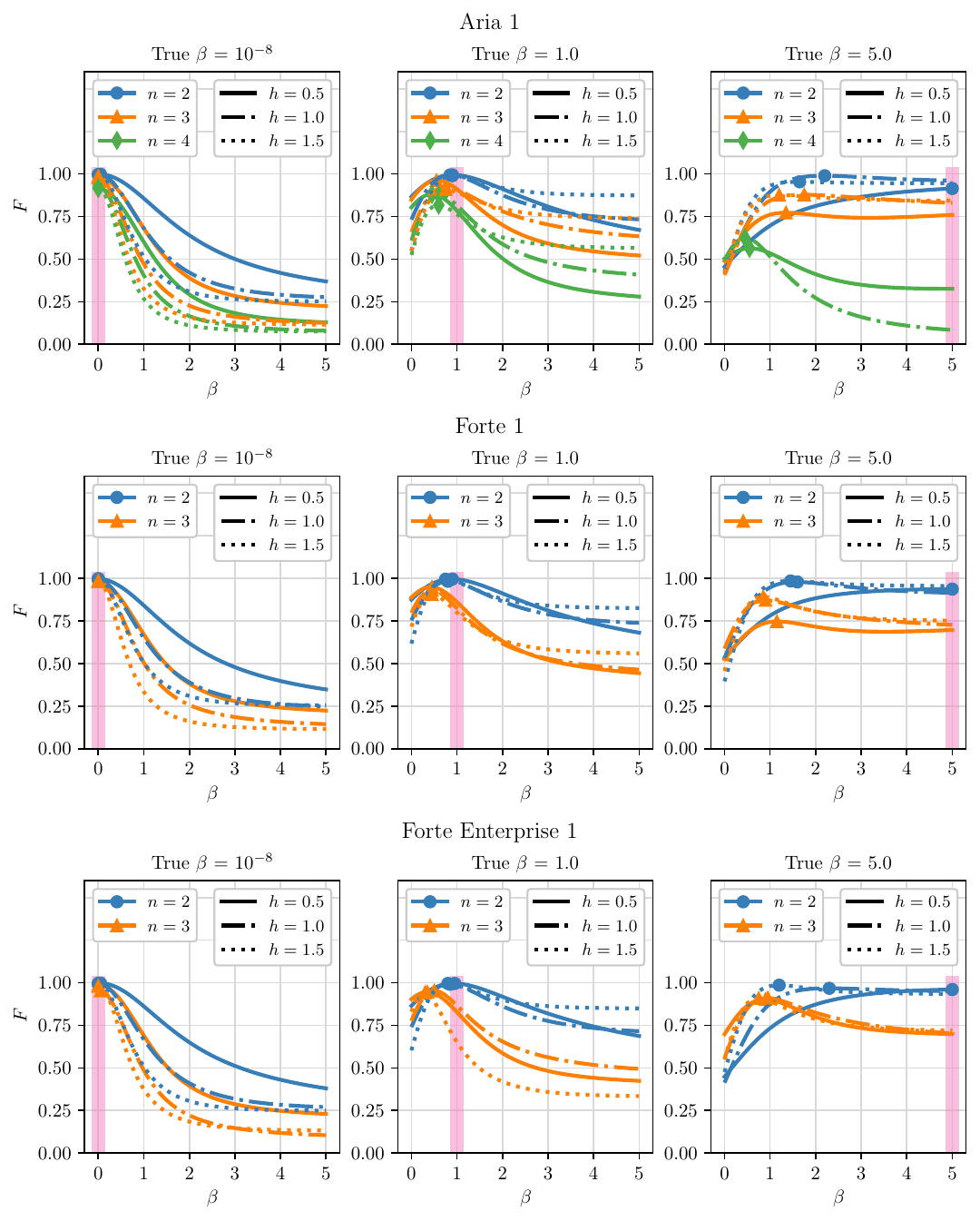}
    \caption{
        Fidelity of a classically prepared Gibbs state with an experimentally prepared Gibbs state, as a function of the $\beta$ used in the classical computation.
        The true value of $\beta$ used in the quantum routine is denoted by the pale vertical bar.
        The $\beta$ that maximizes fidelity for each configuration is denoted by a marker.
        For $\beta=10^{-8}$, fidelity is maximized for the true $\beta$, while as $\beta$ increases, the $\beta$ which maximizes fidelity shifts increasingly far to the left as a function of $n$.
    }
    \label{fig:beta-sweep}
\end{figure*}

As expected from our simulations of Fig.~\ref{fig:simulated-fidelities}, we find that increasing $n$ decreases fidelity across all devices and values of $h$ and $\beta$, and that for $n=2$ the fidelity is near unity across all values of $\beta$. To start, it is interesting to note that the noisy simulator captures the behavior of the real hardware remarkably well. From prior work, we expect the behavior of the simulator to deviate widely from the behavior of the quantum hardware (see, for example, \citet{robertson_simons_2025}), but that is not the case here.

That said, unlike what we see in Fig.~\ref{fig:simulated-fidelities}, increasing $h$ does not always correspond to an increase in fidelity.
More strikingly, while fidelity consistently degrades as $\beta$ increases from $10^{-8}$ to $1.0$, fidelity also typically decreases as we increase $\beta$ to $5.0$.
This is in stark contrast to the noiseless simulations, where the fidelity was high for both high and low values of $\beta$.

From Fig.~\ref{fig:fidelities}, we also see that fidelity does not increase consistently with high values of $\beta$, as expected.
There are two distinct yet related reasons for this.
First, for $\beta=10^{-8}$, the classical optimization loop quickly converged to the minimal cost.
For the larger values of $\beta$, the optimization did not converge, and sometimes terminated relatively far away from the minimal value of the cost function.
This is to say, simulated hardware noise hindered the optimization for large $\beta$.

In addition to this, physical hardware noise present on the device also affected the performance of the algorithm.
Again, the impact of hardware noise on the performance of the algorithm increased with $\beta$.
This can be observed by a parity check on the Gibbs state.
When measured in the $\sigz$ basis, the system ought to exhibit even parity.
Indeed, in $91.30\%$ of our trials, we recovered an even parity through this measurement.
However, the frequency of an even parity result decreases as a function of $n$: it is $95.89\%$ for $n=2$, $88.48\%$ for $n=3$, and $87.35\%$ for $n=4$.

Since noise injects heat into the system, a natural step is to determine how much heating is affecting the algorithm.
More precisely, for which value of $\beta$ will a classically prepared state attain the highest fidelity with our experimental results?
This question is addressed in Fig.~\ref{fig:beta-sweep}.
In this figure, the pale vertical bar denotes the true value of $\beta$, that is, the value of $\beta$ at which we intended to prepare our Gibbs state on the hardware.
The curves denote the fidelity of a classically computed Gibbs state at various values of $\beta$ with the experimental results.
The marker denotes the value of $\beta$ that maximizes fidelity between the classical computation and the hardware results.

First, for $\beta=10^{-8}$, we find that fidelity is maximized between the experimental results and the classical results when the classical results are prepared with the same value of $\beta$.
Thus, we can successfully prepare the Gibbs state on the quantum hardware in the infinite temperature case.
This is intuitive; any thermal fluctuations introduced by the hardware cannot further raise the effective temperature.
As $\beta$ increases, however, we see that the value of $\beta$ that maximizes fidelity between the classical algorithm and the quantum algorithm shifts to the left.
Moreover, this shift increases consistently as $n$ increases.
Again, this is intuitive; thermal fluctuations introduced by the hardware effectively increase the temperature of the prepared state.
This temperature increase is more dramatic for lower initial temperatures (higher values of $\beta$).
Moreover, larger systems (larger $n$) allow for increased opportunity for hardware noise to play a role.
What is less intuitive, however, is that the shift away from the true value of $\beta$ is inversely correlated with $h$---when $n=2$ and $h=0.5$, fidelity is maximized by the true value of $\beta$, but it is maximized for much lower values of $\beta$ for the larger values of $h$.
This contrasts with the results from the noiseless simulator, where increasing $h$ increased the overall fidelity.

\begin{figure}
    \centering
    \includegraphics[width=\linewidth]{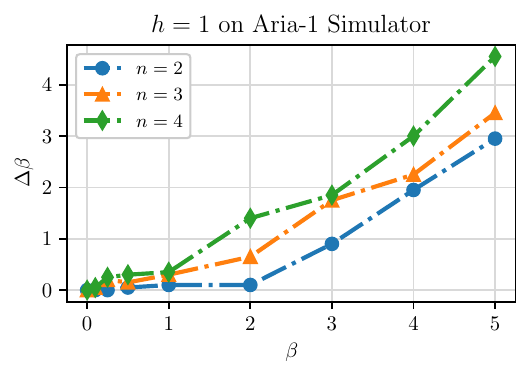}
    \caption{
        The change in $\beta$ required in the classical Gibbs state preparation to maximize fidelity with simulated results executed at various true values of $\beta$.
        All results in this plot are obtained from the IonQ Aria-1 simulator.
        Once again, we see $\Delta\beta$ increase as both a function of $\beta$ and of $n$.
    }
    \label{fig:delta-beta}
\end{figure}

Thus, from Fig.~\ref{fig:beta-sweep}, we see that outside of the infinite temperature case, the quantum hardware consistently overshoots the temperature of the prepared Gibbs state.
To probe this discrepancy, we simulated the GSP routine for several values of $\beta$ on the IonQ Aria-1 simulator\footnote{Due to the remarkable agreement between the hardware and the simulator for this algorithm, we are confident in the results derived from the simulator.} and evaluated the change in $\beta$, $\Delta\beta$, introduced in the classical Gibbs state preparation routine to maximize fidelity with the simulated result.
The results are plotted in Fig.~\ref{fig:delta-beta}.
These results agree with those of Fig.~\ref{fig:beta-sweep}; once again, $\Delta\beta$ increases with both $\beta$ and $n$.

\section{Concluding Remarks}\label{sec:conclusion}

In this paper, we benchmarked the performance of a hybrid quantum--classical GSP routine on trapped-ion quantum computers.
We prepared Gibbs states for the TFIM Hamiltonian of Eq.~\ref{eq:ising-hamiltonian}, and we varied $n\in\{2,3,4\}$, $h\in\{0.5,1.0,1.5\}$, and $\beta\in\{10^{-8},1,5\}$ across our experiments.
On a noiseless simulator, we found that the GSP routine worked exactly for the smallest $n$ and the extreme values of $\beta$.
For intermediate values of $\beta$, fidelity decreased, reaching a minimum near $\beta\sim1$.

On real hardware, the GSP routine maintained high (though not perfect) fidelity for $n=2$ and $\beta=10^{-8}$.
Similarly, fidelity decreased for $\beta=1$.
However, in sharp contrast to the noiseless simulator, fidelity also dropped dramatically for $\beta=5$.
This decrease in fidelity is due to hardware noise, which digitally heats the system above the desired temperature.
We probed this discrepancy and found that this change in temperature between the desired Gibbs state and the prepared Gibbs state increases as a function of $\beta$.
This result is of significant importance to researchers who seek to prepare Gibbs states as a preliminary step for a practical quantum algorithm---the hardware must be carefully characterized beforehand to determine the effective temperature increase due to noise for the desired Gibbs state.

Second, contrary to expectations, it was discovered that the noisy simulator performed very accurately for this GSP routine.
Work should be done to identify the applications in which simulator predictions can be trusted.

\begin{acknowledgments}

M.C. acknowledges funding from Project QAMALA (Quantum Algorithms and MAchine LeArning) financed by the Maltese Ministry for Education, Sport, Youth, Research, and Innovation (MEYR) Grant ``2025301 UM MinED''. S.D. acknowledges support from the John Templeton Foundation under Grant No. 63626. E.D. acknowledges U.S. NSF under Grant No. OSI-2328774. This work was supported by the U.S. Department of Energy, Office of Basic Energy Sciences, Quantum Information Science program in Chemical Sciences, Geosciences, and Biosciences, under Award No. DE-SC0025997.

\end{acknowledgments}

\begin{table}[bp]
    \centering
    \begin{tabular}{|c|c|c|c|}
        \hline
        Parameter & Aria 1 & Forte 1 & Forte Enterprise 1 \\
        \hline
        \hline
        T1 Time & 100 s & 100 s & 188 s \\
        \hline
        T2 Time & 1 s & 1 s & 0.95 s \\
        \hline
        1-q Gate Speed & 135 $\mu$s & 130 $\mu$s & 63 $\mu$s \\
        \hline
        2-q Gate Speed & 600 $\mu$s & 970 $\mu$s & 650 $\mu$s \\
        \hline
        Readout Speed & 300 $\mu$s & 150 $\mu$s & 250 $\mu$s \\
        \hline
        Reset Speed & 20 $\mu$s & 50 $\mu$s & 150 $\mu$s \\
        \hline
        1-q Gate Fid. & 0.9998 & 0.9998 & 0.9998 \\
        \hline
        2-q Gate Fid. & 0.9799 & 0.9849 & 0.9915 \\
        \hline
        SPAM Fid. & 0.9951 & 0.9946 & 0.9939 \\
        \hline
        \# Qubits & 25 & 36 & 36 \\
        \hline
        2-q Gate & \texttt{MS} & \texttt{ZZ} & \texttt{ZZ} \\
        \hline
        \hline
        \# Circuits & 972 & 405 & 324 \\
        \hline
        Cost (USD) & \$72,433.44 & \$53,046.90 & \$41,437.52 \\
        \hline
    \end{tabular}
    \caption{
        The parameters for the IonQ hardware at the time of job submission.
        The \textit{\# Circuits} parameter denotes the number of circuits executed on each device during the course of these experiments.
        Similarly, the \textit{Cost (USD)} parameter denotes how much money (USD) was spent on compute credits for each device.
    }
    \label{tab:hardware}
\end{table}

\subparagraph{Data availability}

All data and source code files for the project are available at \url{https://github.com/reecejrobertson/GibbsStatePreparation}.

\appendix

\begin{figure*}[t]
    \centering
    \includegraphics[width=\textwidth]{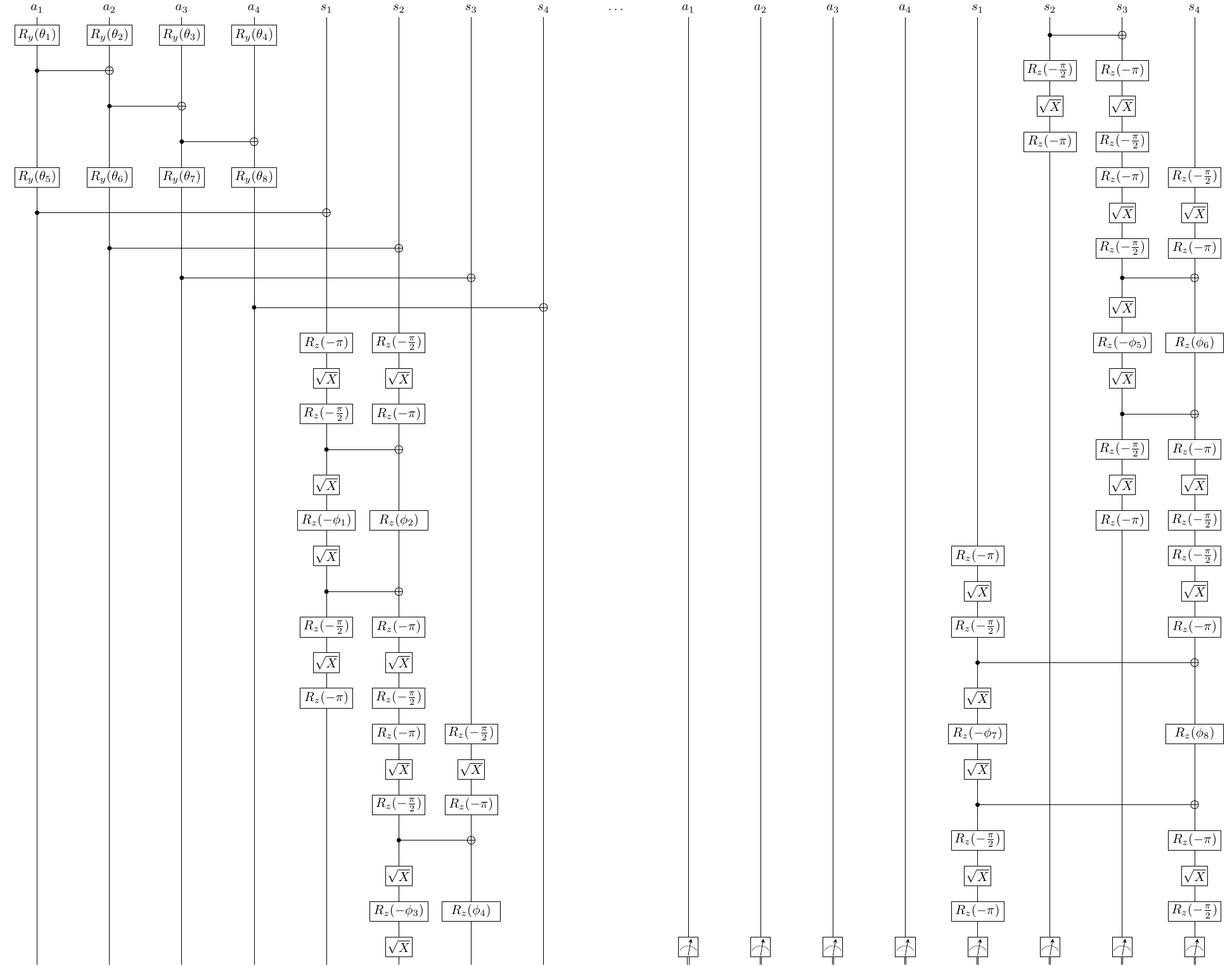}
    \caption{
        The complete GSP algorithm for a 4-qubit system.
        Time advances vertically down each wire, and the eight wires on the left precede the eight wires on the right in time.
        All qubits are initialized in the $|0\rangle$ state.
    }
    \label{fig:alg-example}
\end{figure*}

\section{Hardware Parameters}\label{sec:hardware}

Table~\ref{tab:hardware} details the hardware parameters of the three IonQ devices at the time of job submission.
All three devices have an all-to-all topology, and the native gate set includes the \texttt{GPi} (general $\pi$ rotation), \texttt{GPi2} (general $\pi/2$ rotation), and \texttt{Virtual Z} single-qubit gates in addition to the two-qubit gate given in the table.
Each of the IonQ native gates is detailed in IonQ's documentation \cite{ionq_docs}.

\section{Complete Algorithm Example}\label{sec:example}

Fig.~\ref{fig:alg-example} depicts a complete example of the GSP algorithm for a four-qubit system.
One ancilla layer and one system layer are used, which is representative of the largest problem that was executed on the IonQ hardware.

There are some things to be said about the final measurement operation shown at the end of the circuit.
First, if one is using the GSP algorithm as a subroutine for a larger quantum operation, then the measurement of the system register should be omitted.

Second, although Fig.~\ref{fig:alg-example} presents the measurement of the ancilla register at the end of the algorithm, this measurement can be performed much earlier.
As explained by \citet{consiglio_variational_2024}, the transversal layer of \texttt{CNOT} gates can be replaced by a midcircuit measurement of the ancilla register, and then a layer of $\sigx$ gates on the system qubits conditioned on the measurement result.
This has the advantage of allowing the reuse of the ancilla register for the system register, halving the number of qubits required for the algorithm.
However, such a procedure requires that the hardware support midcircuit measurement, which was not the case for the IonQ devices tested at the time of execution.
Alternatively, recalling that by the deferred measurement principle, measurements commute with controls \cite{nielsen_quantum_2010}, the ancilla register could be measured immediately after $U_A(\boldsymbol\theta)$ (before the transversal \texttt{CNOT}s).

Finally, we use the same measurement symbol on the ancilla and the system registers.
Throughout the GSP routine, the measurement on the ancilla register is always taken in the $\sigz$ basis (indeed, it is from this value that we compute the von Neumann entropy of Eq.~\eqref{eq:cost-function}).
The measurement on the system register, however, varies depending on the context.
In the parameter training protocol, the system is measured in both the $\sigz$ and $\sigx$ bases.
Both are required to compute Eq. \eqref{eq:cost-function}.
When we executed the GSP algorithm on real hardware, on the other hand, we performed quantum state tomography on the system register.
This allowed us to reconstruct the density matrix of the system register and thus evaluate the fidelity of the prepared Gibbs state.

\bibliography{bib}

\end{document}